# Asymmetric fitting function for condensed-phase Raman spectroscopy

Vitaly I. Korepanov*, Darya M. Sedlovets

Asymmetric lineshapes are experimentally observed in Raman spectra of different classes of condensed matter. Determination of the peak parameters, typically done with symmetric pseudo-Voigt functions, in such situations yields unreliable results. While wide choice of asymmetric fitting functions is possible, for the function to be practically useful, it should satisfy several criteria: simple analytic form, minimum of parameters, description of the symmetric shape as "zero case", estimation of the desired peak parameters in a straightforward way and, above all, adequate description of the experimental data. In this work we formulate the asymmetric pseudo-Voigt function by damped perturbation of the original symmetric shapes with one asymmetry-related parameter. The damped character of the perturbation ensures by construction the consistent behavior of the line tails. We test the asymmetric function by fitting the experimental Raman spectra. The results show that the function is able to describe a wide range of experimentally observed asymmetries for different nature of asymmetric broadening, including 3D and 2D crystals, nanoparticles, polymer, molecular solid and liquid.

## Introduction

Analysis of big sets of spectral data requires fast and precise estimation of peak parameters, such as position of the maximum, full width at half maximum (FWHM) and peak area. This can be done in a straightforward way for symmetric line shapes by fitting the spectra with Lorentzian, Gaussian or Voigt functions. However, for solid state Raman spectroscopy, a variety of local environments lead to inhomogeneous broadening, and as a consequence to asymmetric line shape. This effect is often observed for nanoscale materials, and can have multiple origins, such as surface states[1–3], phonon confinement by grain boundaries[4,5], anharmonicity[6], local heating by the laser[7], stoichiometry variations and lattice defects. In liquid samples, the asymmetry is also observed, and the peak broadening is attributed to local "microenvironments"[8]. For some particular cases (e.g. phonon confinement), physical models are available to describe the asymmetric line shapes, but for the general case the inhomogeneous broadening can hardly be defined in terms of equations.

Regardless of the nature of broadening for the particular sample, it is often desirable to fit the peaks with a simple analytic function to estimate its parameters[8,9]. This is especially useful for Raman mapping, in which peak parameters at each point of the map can correlate to certain physical property of the sample, such as concentrations of the components, thickness in case of the thin film, crystallinity, lattice strain etc.

In the present work, we construct the asymmetric fitting function by a perturbation of the pseudo-Voigt profile. Only one additional parameter is required in this approach. We test the applicability of the function by fitting the asymmetric experimental lines of several fundamentally different classes of samples, including 3D and 2D crystalline solids, nanoparticles, polymer, molecular solid and liquid.

## Formulation of the asymmetric lineshape

Many choices of asymmetric line profiles are possible[8–10]. However, for the fitting function to be practically useful, it must have a simple analytic form with minimum parameters, include symmetric shape as "zero case", yield the desired peak parameters in a straightforward way and, most importantly, adequately describe the experimental data. A convenient way is to modify the widely-used pseudo-Voigt function. As one of the options, this can be done by introducing the variable (wavenumber-dependent) FWHM ($\Gamma$). In the work[8], $\Gamma$ was varied sigmoidally:

$$\Gamma(\omega) = \frac{2\Gamma_0}{1+\exp[a(\omega-\omega_0)]}, \qquad (1)$$

where $a$ is the asymmetry parameter, and $\omega_0$ is the peak position. Next step, $\Gamma$ in the Lorentzian and Gaussian lineshapes is replaced by $\Gamma(\omega)$ from eq.(1). This approach was shown to adequately reproduce the experimental asymmetric lineshape in IR spectra, however, it has one major drawback. The maximum position of the resulting pseudo-Voigt function does not coincide with $\omega_0$ (fig. 1). Another problem is that FWHM is also not derived from the fit in a convenient way.

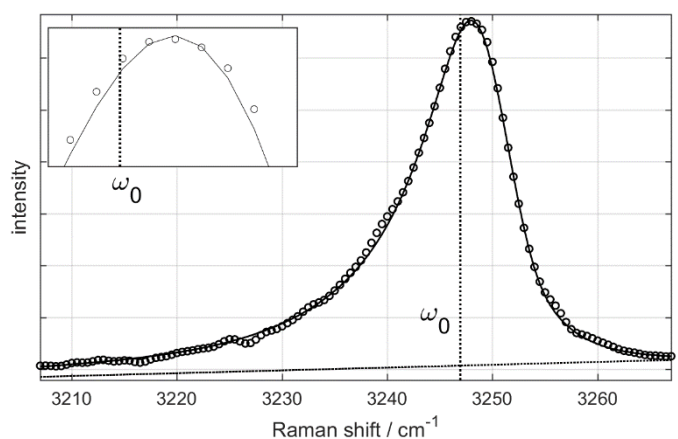

Figure 1. Numerical artefact of asymmetric fitting function with sigmoidal $\Gamma$ (eq. 1): $\omega_0$ does not coincide with the position of the maximum. Experimental spectra of HOPG (circles) and fit with asymmetric pseudo-Voigt function from ref.[8]. Dotted line is the position of the resulting $\omega_0$.

For this reason, it is not convenient to use such fitting function for finding the bandshape parameters. In the present work, we aim at defining a simple asymmetric function, the maximum of which would exactly coincide with $\omega_0$ of the parent symmetric function.

We take the Lorentzian and Gaussian shapes normalized to unit area, with the same FWHM ($\Gamma$). The pseudo-Voigt function is a linear combination of these two profiles[8].

$$Lor(\omega) = \frac{A}{2\pi} * \frac{\Gamma}{(\omega-\omega_0)^2 - (\Gamma/2)^2} \quad (2)$$

$$Gauss(\omega) = \frac{A}{\Gamma} * \sqrt{\frac{4\ln 2}{\pi}} \exp\left[-4\ln 2 \left(\frac{\omega-\omega_0}{\Gamma}\right)^2\right] \quad (3)$$

We introduce a perturbation of the form of damped sigmoidal shape:

$$p(\omega) = 1 - a * \frac{\omega-\omega_0}{\Gamma} * \exp\left[-\frac{(\omega-\omega_0)^2}{2*(2\Gamma)^2}\right] \quad (4),$$

where $a$ is the asymmetry parameter. The Gauss-like function has the dispersion twice that of the original pseudo-Voigt profile; although different dispersions are possible, according to our tests, the 2$\Gamma$ shows a good description of experimental data. Unlike commonly used sigmoidal shape[9,11], the damped perturbation keeps the tails of the asymmetric profile close to those of the parent Gaussian and Lorentzian (fig. 2).

The pseudo-Voigt profile is constructed with the wavenumber $\omega$ replaced by $\omega*p(\omega)$:

$$f(\omega, a) = m * Gauss(\omega * p(\omega)) + (1-m) * Lor(\omega * p(\omega)) \quad (5)$$

The resulting asymmetric line profile has one parameter to control asymmetry ($a$). The sign of asymmetry parameter decides to which side of the central frequency "tailing" occurs: for $a=0$ it coincides with the symmetric (un-perturbed) function, for $a<0$ it has a skew towards low wavenumbers, and for $a>0$ the skew is positive.

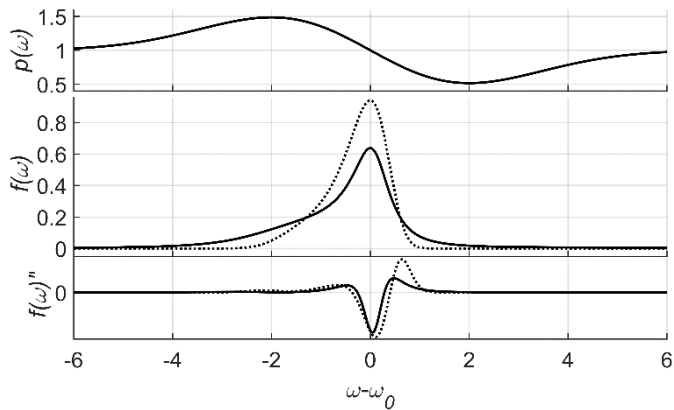

Figure 2. Perturbation $p(\omega)$ (top), asymmetric line shapes $f(\omega)$ for Lorentzian (middle, solid line) and Gaussian (middle, dotted line) and the corresponding second derivatives $f''(\omega)$ (bottom). Asymmetry parameter $a$ is set to -0.4, $\Gamma$=1.

The conventional criterion to assign the spectral feature to a single band is the detection of only one minimum of the second derivative[12], while two minima indicate the presence of two bands. According to this formulation, we can estimate the physically meaningful values of the asymmetry parameter by plotting $f(\omega)$ for different $a$. From such estimation, the asymmetric profile $f(\omega)$ corresponds to a single band approximately within $a$ = [-0.4…0.4], while outside this range the second minimum appears. Within this interval, the line shape varies smoothly with $a$.

Unlike the band position $\omega_0$, FWHMs and areas of the perturbed lineshapes don't coincide with those the original symmetric functions (except for the case $a$=0). No exact analytic equation is available to describe the parameters of the perturbed peaks, however, an approximate dependence can be derived from numerical analysis in the following way: first, peak parameters are estimated for the whole range of $a$; next, the results are fit with the polynomial. The procedure is illustrated on fig. 3 for the peak width. For both Gaussian and Lorentzian shapes, FWHM has the same dependence:

$$FWHM(a) = \Gamma_0 * (1 + 0.40a^2 + 1.35a^4) \quad (6)$$

Same estimation can be done for the peak areas. For the latter, the coefficients are different for Gaussian and Lorentzian shapes:

$$Area(Lor, a) = Area(Lor, 0) * (1 + 0.69a^2 + 1.21a^4) \quad (7)$$

$$Area(Gauss, a) = Area(Gauss, 0) * (1 + 0.67a^2 + 3.43a^4) \quad (8)$$

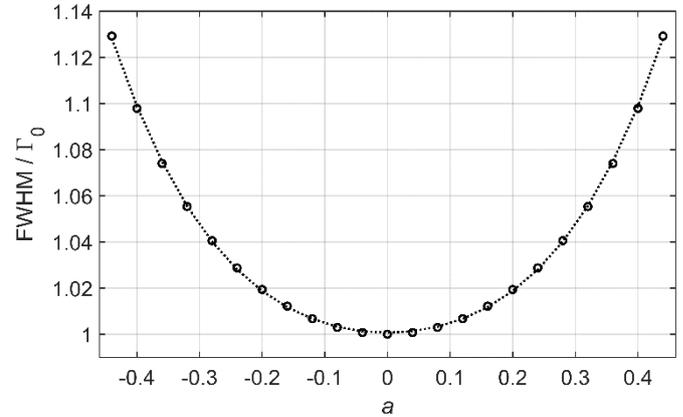

Figure 3. Dependence of the FWHM of the asymmetric peak on the parameter $a$. Circles: values calculated from eq. 5, dotted line: polynomial fit (eq. 6).

We should emphasize that the damped perturbation $p(\omega)$ introduced here is a purely mathematical step, and does not rely on a physical model. The purpose of the following section is to test the fitting function on the systems, for which the physical nature of the asymmetry is different.

## Fitting of experimental spectra

We tested the asymmetric fitting function on several samples of different nature and different origins of asymmetry. For each spectrum, the fit was done with symmetric ($a=0$) and asymmetric fitting function, and peak parameters were estimated from the fit. The experimental area of the bands was found by numerical integration within the fitting region. The results are summarized in Table 1 and plotted in the figures below.

**3D crystal (ZnO).** Classic example of asymmetric line shape in Raman spectra of crystals is the $E_2^{high}$ mode of ZnO, the skew of which is attributed to the effect of anharmonicity[6]. We study the spectral pattern, detected from single crystal in $x(yy)x$ geometry (Porto notation). Two peaks are overlapping in this spectral region, for which the asymmetries were found to be different: the $E_2^{high}$ mode has pronounced asymmetry ($a=-0.22$), while the $E_1$ mode is close to symmetrical shape ($a=0.06$). As can be seen from fig. 4, asymmetric functions give significantly better description of the spectral shapes. The differences in peak positions are about 0.3 cm$^{-1}$ with the asymmetric function giving better estimation of the maximum; FWHMs are close for both fittings, because the main difference is in the tails for the lineshapes. The areas were also found to be close to each other, however, taking into account the difference in the baseline for the two fits, the asymmetric fit overall gives significantly better description of the experimental spectral shape.

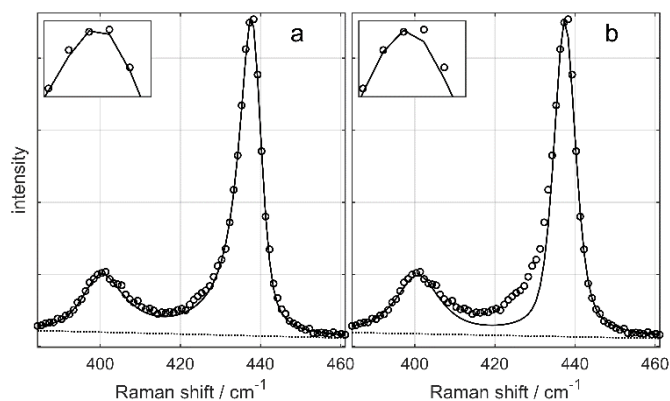

Figure 4. Experimental spectra of ZnO (circles) and fit with asymmetric (a) and symmetric (b) functions. Dotted line is linear baseline. Inset shows the close-up of the main peak maximum.

**2D crystal: HOPG.** For graphene and related systems, asymmetry is of high importance, because, for example, the G band shows a shift depending on lattice stress, local disorder and doping[13–15]. Therefore, for non-uniform samples the bands often have certain asymmetry. Higher-order bands are asymmetric even for such ordered systems as HOPG[16]. The origin of the bandshape asymmetry is not discussed in the literature. In this work we consider the 2D' band of HOPG, which is a perfect model for testing, because it is an isolated spectral line with pronounced asymmetry, which has been reported before and can be easily reproduced experimentally[17,18]. As can be seen from fig. 5, the asymmetric fit shows much better description of the experimental data for both peak position and width.

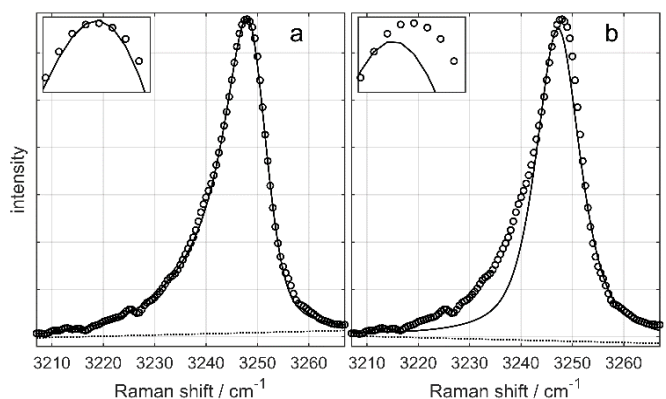

Figure 5. Experimental spectra of HOPG (circles) and fit with asymmetric (a) and symmetric (b) functions. Dotted line is linear baseline. Inset shows the close-up of the main peak maximum.

**Nanoparticles.** Strictly speaking, Raman spectral patterns of nanoparticles cannot be interpreted as individual lines. More accurately, they belong to a superposition of bands from different $k$-points of Brillouin zone[5], and this can be more complex in case if the sample is not monodisperse[19]. Nevertheless, for quantitative analysis it is highly desirable to have some fitting function, which would be able to adequately describe the observed line shapes. In this work we studied the previously reported experimental spectrum from diamond nanoparticles of 2.7 nm median size, which has an isolated line of diamond phonon[20]. The fit results are plotted in fig. 6. The asymmetry parameter from the fit was found to be $a=-0.43$, which is outside the approximate "single-band" range defined above. We should however note that, due to the complex origin of this line[5], the single-minimum criterion for the second derivative[12] is not applicable here.

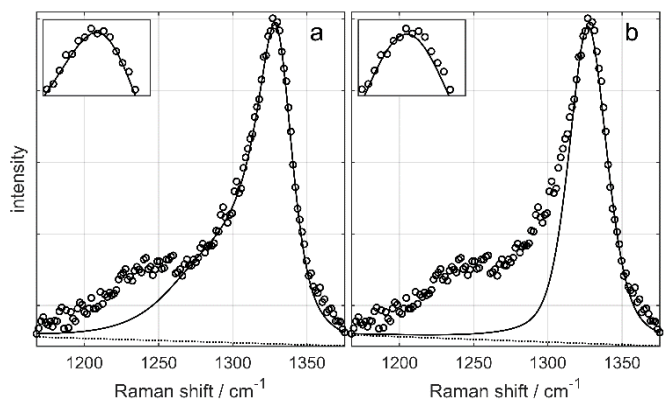

Figure 6. Experimental spectra of ND (circles) and fit with asymmetric (a) and symmetric (b) functions. Dotted line is linear baseline. Inset shows the close-up of the main peak maximum.

**Polymers** represent another example of partially disordered system. Depending on the degree of crystallinity and the history of the sample, a variety of atomic environments can be observed[21]. In this work, we studied PTFE, a system of high practical importance for spectroscopy. The peaks of this polymer at 732 and 1379.5 cm$^{-1}$ are used as spectroscopic standard for Raman shift calibration[22,23]. In this work, we take the CF$_4$ stretching band at 732 cm$^{-1}$ which has a pronounced asymmetry[23]. The peak positions defined by symmetric and asymmetric functions show 0.2 cm$^{-1}$ difference (table 1), with significantly better fit for the latter one (fig. 7).

the asymmetric fitting function works well for this line (fig. 9), yielding good estimation of the peak parameters (table 1).

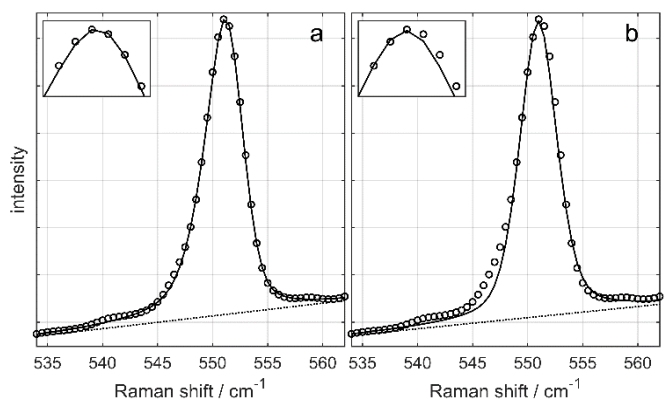

Figure 8. Experimental spectra of PMTN (circles) and fit with asymmetric (a) and symmetric (b) functions. Dotted line is linear baseline. Inset shows the close-up of the main peak maximum.

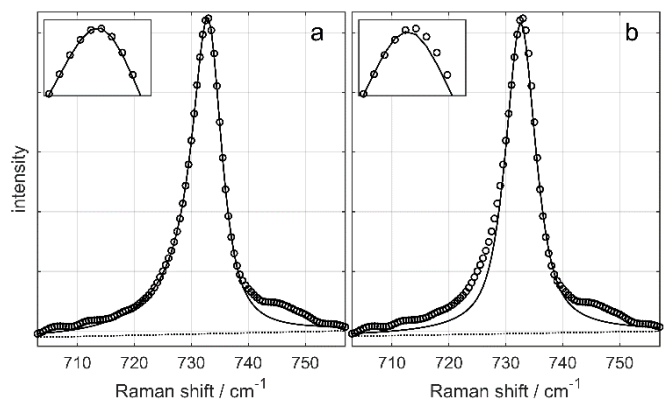

Figure 7. Experimental spectra of PTFE (circles) and fit with asymmetric (a) and symmetric (b) functions. Dotted line is linear baseline. Inset shows the close-up of the main peak maximum.

**Molecular solids** also can give asymmetric Raman bands. In this work, we take the 551 cm$^{-1}$ band of pyromellitic tetranitrile (PMTN), a practically valuable precursor for 2D polymers[24]. The origins of the asymmetry may include grain boundaries different phases and other reasons. The asymmetry of the peak is moderate (fig. 8), which corresponds to the asymmetry parameter of $a$=-0.17.

**Liquids,** due to rotational averaging, typically don't display inhomogeneous broadening of the Raman bands. Nevertheless, for some cases, asymmetric lines can be observed. One common example is the C=O band of acetone, which has a pronounced asymmetry[25]. In the present work, we show that

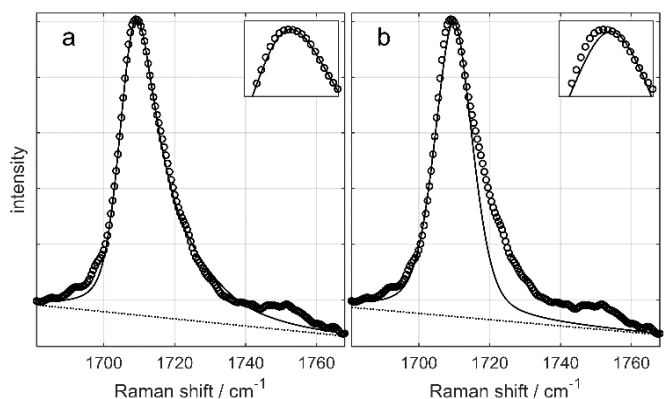

Figure 9. Experimental spectra of acetone (circles) and fit with asymmetric (a) and symmetric (b) functions. Dotted line is linear baseline. Inset shows the close-up of the main peak maximum.

**Table 1**. Peak parameters estimated from the fitting with asymmetric and symmetric fitting functions. Peak positions ($\omega_0$) and FWHM in cm$^{-1}$, other parameters dimensionless.

|  | ZnO | HOPG | ND | PTFE | PMTN | acetone |
|---|---|---|---|---|---|---|
| $\omega_0$ asym. | 437.66 / 400.44 | 3247.90 | 1328.30 | 732.82 | 551.15 | 1709.34 |
| $\omega_0$ sym. | 437.39 / 400.37 | 3247.24 | 1327.12 | 732.62 | 550.96 | 1709.83 |
| asymmetry ($a$) | -0.22 / 0.06 | -0.34 | -0.43 | -0.20 | -0.17 | 0.35 |
| FWHM asym. | 6.63 / 11.70 | 10.81 | 36.27 | 6.49 | 4.10 | 13.98 |
| FWHM sym. | 6.33 / 11.44 | 9.81 | 30.57 | 6.22 | 3.91 | 12.29 |
| area asym./exp. | 0.91 | 0.94 | 0.84 | 0.92 | 0.95 | 0.89 |
| area sym./exp. | 0.89 | 0.79 | 0.58 | 0.81 | 0.90 | 0.72 |
| m asym. | 0.19 / 0.23 | 0.32 | 0.12 | 0.21 | 0.50 | 0.40 |
| m sym. | 0.19 / 0.22 | 0.33 | 0.44 | 0.08 | 0.50 | 0.50 |

## Experimental details

For ZnO, the sample was 10x10x0.5 mm single crystal from Semiconductor Wafer Inc, the measurement was done at NCTU (Taiwan) with 532 nm excitation with the lab-built Raman system described elsewhere[26]. Accumulation time was 4x30s. The Raman shift was calibrated with Ne lamp. The ND spectrum was measured with 355 nm excitation; the sample history and experimental details were reported before[20].

For the samples of HOPG (Sigma-Aldrich), PMTN (97% mass $C_6H_2(CN)_4$ by Sigma Aldrich), acetone (99.8% $C_3H_6O$ by EKOS-1), Raman spectra were measured in the 300-3700 $cm^{-1}$ range with a Bruker Senterra micro-Raman system under 532 nm excitation. The laser power was 20 mW. Typically, 2 accumulations for 15 s were done for each spectrum, excluding HOPG (5 accumulations for 30 s).

Fitting of the experimental data with equation 5 was done as a least-square minimization, with $\omega_0$, $\Gamma_0$, asymmetry, amplitude, and linear baseline as fitting parameters. For fitting with the asymmetric function, resulting FWHMs and peak areas were estimated from equations 6-8.

## Conclusions

Asymmetric pseudo-Voigt function defined in this work has a simple analytic form with only one additional parameter (*a*) describing the degree of asymmetry. It smoothly changes with *a*, and has a good flexibility, describing a wide range of possible lineshapes. This fitting function can be efficiently used to extract peak parameters (peak position, FWHM and band area) from the experimental data. By the construction, the function has a good asymptotic behaviour of the tails within the physically meaningful range of *a*, which is guaranteed by the damped character of perturbation. Moreover, this also allows one to use the fitting function to estimate the area in case of partially overlapping peaks.

We tested the function by fitting the asymmetric experimental lines of several fundamentally different classes of samples, including 3D and 2D crystalline solids, nanoparticles, polymer, molecular solid and liquid. For all cases, the asymmetric function was found to give a good description of the experimental data.

## Acknowledgements

We acknowledge Prof. Igor G. Korepanov (Moscow Aviation Institute) for discussion of mathematical aspects of this work.